\documentstyle[11pt]{article}

\newcommand{\slashed}[1]%
{{%
\setbox0=\hbox{$#1$}%
\makebox[0pt][l]{$#1$}%
\makebox[\wd0][c]{/}%
}}

\title{Self-consistent Study on Color Transport
       in the Quark Gluon Plasma at
       Finite Chemical Potential}
\author{Hou Defu$^\dagger$, Stefan Ochs$^{\ast\dagger}$, Li Jiarong$^\dagger$}

\begin{document}

\maketitle

\begin{center}
 {\small $^\dagger$Institute of Particle Physics,
 Hua-Zhong Normal University\\
 Wuhan 430070, China\\
 
 $^\ast$Institut f\"ur theoretische Physik,
 93040 Uni Regensburg, Germany }
\end{center}

\begin{abstract}
 We calculate the relaxation time self-consistently
 to study the damping of collective color modes and
 the color conductivity in a QGP by deriving self-consistent
 equations for the damping rates of gluons and quarks
 to leading order QCD by TFD
 including a chemical potential for quarks. We show
 that the damping rates are not sensitive to the chemical
 potential whereas color conductivity is enhanced considerably.
\end{abstract}

\section{Introduction}

Heavy ion experiments at RHIC and SPS have now reached a stage
where we can expect to probe QCD matter beyond its hadronic state
in the quark gluon plasma (QGP) phase.
For this phase of essentially free colored particles we expect
a behaviour similar to the one in a QED plasma, that is
Debye screening of color electric charges and their interactions,
as well as collective color modes \cite{Muller},\cite{Heinz}.

Collective behaviour might be observable in ultrarelativistic
heavy ion collisions as long as the lifetime of the QGP exceeds
the time scale  for dissipative processes. Therefore it is
important to calculate the damping rate for collective phenomena.

So far, mostly momentum relaxation rates were considered
\cite{Dan},\cite{Baym}, which allow then estimates for viscosity,
thermal conductivity and energy transport in the plasma.
Some work has been done also on color relaxation
\cite{Burgess},\cite{Gullasy},\cite{Thoma} to calculate the
color conductivity and the color diffusion coefficients.

We want to extend that work and calculate the static color
conductivity $\sigma$ including a chemical potential for quarks. The
chemical potential has been estimated to become zero only when nuclear
transparency sets in and reach values up to 1 -- 2 times the
temperature in the region where stopping occurs \cite{Viji}.

The (infrared) divergence occuring in the calculation of color conductivity,
which can be related to long range (color) magnetic interactions,
will be treated in our approach self-consistently by including
a finite width $\nu$ for quarks and gluons which acts as a
self-consistent damping term.
In previous works \cite{Burgess},\cite{Gullasy},\cite{Thoma}
this divergence was regularized  using a (non-perturbative)
magnetic screening mass $m_{g}$ as an additional cutoff parameter.
We will show explicitely, that a finite width will regularize the
divergence self-consistently.

The paper will be organized as follows:
In section \ref{sec2} we will get the self-consistent equations
for the width $\nu$ of both gluons and quarks
to first order perturbation theory in the
hard thermal loop approximation using the real time formalism
of thermal field dynamics (TDF) including a chemical potential
$\mu$ for quarks.
In section \ref{sec3} we will discuss the self-consistent
(numerical) solutions of $\nu$ and $\sigma$ as functions of
temperature and chemical potential.
We will summarize our results in section \ref{sec4}.

\section{Damping rates of fast moving particles in a QGP}
\label{sec2}

If one considers a fast moving particle with momentum scale
$p\gg g T$ going through a weakly coupled QGP ($g\ll 1$),
the relativistic dispersion relation for the particle, the
transverse gluon for instance, can be written according to
the poles of its propagator in the  following form
\begin{equation}\label{dispersion}
 E(k) = \omega(k) - i \nu_T,
\end{equation}
where $\nu_T$ is the damping rate of the transverse gluon.

It is sufficient to consider only the fast moving particles since for the
static color conductivity the momentum exchange is small (\cal{O}$(gT)$). 
This results in a damping rate for the slow moving particles of order $g^2
T$ in the weak coupling limit, which is small compared to the one for the fast
moving particles of order $g^2 T/\ln (1/g)$ \cite{Braaten90}.
This corresponds to the scenario of a realistic hot QGP, where the momentum of
a particle should be of order $T$, that is hard ($\langle p \rangle = 3T$), but
where the momentum exchange can be rather small.
In addition, the color diffusion coefficient $d_c = p \cdot u \ \nu_T$ is
proportional to the momentum of the particle 
($u_\mu$ being the four velocity in the ensemble rest frame), 
so that color diffusion is suppressed for slow particles. 

In covariant gauge the gluon propagator is
\begin{equation}\label{propagator}
 G_{\mu\nu} (k) = A_{\mu\nu} (k) \Delta_T (k) +
 B_{\mu\nu} (k) \Delta_L (k) +
 D_{\mu\nu} (k) \Delta_\xi (k).
\end{equation}
$\Delta_T$ and $\Delta_L$ are the transverse and longitudinal
gluon propagators respectively, given by
\begin{equation}\label{gluonprop}
 \Delta_{T,L} = {1\over k^2 - \Pi_{T,L}},
 \end{equation}
$\xi$ is the gauge parameter with associated propagator
\begin{equation}\label{gaugeprop}
 \Delta_\xi = \xi {1 \over k^2}
\end{equation}
and $A_{\mu\nu}, B_{\mu\nu}, D_{\mu\nu}$ are functions of
the loop momentum $k$,
$k=(k_0, K)$,
\begin{eqnarray}\label{ABDfunc}
 A_{\mu\nu}(k) &=& \delta_{\mu i} \left( \delta_{ij} -
 {k_i k_j\over K^2} \right) \delta_{j\nu},
 \nonumber\\
 B_{\mu\nu} (k) &=& \left( \delta_{\mu 0} -
 {k_{\mu} k_0\over k^2} \right){k^2\over K^2}
 \left( \delta_{\nu 0} - {k_{\nu}k_0\over k^2} \right),
 \nonumber\\
 D_{\mu\nu} (k) &=& {k_\mu k_\nu\over k^2}.
\end{eqnarray}
$\Pi$ is the polarization tensor, and the dispersion relation
for the transverse gluon can be cast into the form
\cite{Burgess}
\begin{equation}\label{disp-trans-gluon}
 E^2 - P^2 = {1\over 2}
 \left(
 \Pi_i^i - {p^i p^j \Pi_{ij} \over P^2}
 \right) \equiv F.
\end{equation}
From Eqs.(\ref{dispersion}) and (\ref{disp-trans-gluon})
one obtains the damping rate for a transverse gluon on the
mass shell \cite{Burgess}:
\begin{equation}\label{gluondamp}
 \nu_T = -{\mbox{Im} F\over 2\vert P\vert}.
\end{equation}
The crucial step in evaluating Eq.(\ref{gluondamp}) will be
to calculate the imaginary part of the polarization tensor.
The one-loop graphs for $\Pi$ in $SU(N)$---QCD are given in Fig.\ref{fig.1}.

Since the imaginary part of the tadpole diagram vanishes, it
does not contribute to the damping rate. For a fast moving
particle with hard external momenta it is sufficient to use
the effective propagator for one of the internal lines and bare
propagators and vertices otherwise according to the resummation
scheme of Braaten and Pisarski \cite{Braaten90},\cite{Pisarski}.
Because of the Bose enhancement factor $1 + n_B$ for the gluon
loop, the diagram of Fig.\ref{fig.1}(a) will give the leading order
contribution to the imaginary part of the polarization tensor,
whereas the quark and ghost loop can be neglected due to the Pauli
blocking factor $1 - n_f$ and will contribute only to higher order
in the coupling constant.
Therefore up to leading order we only have to calculate the
three-gluon one-loop resummation graph of Fig.\ref{fig.1}(a).

In the static limit the contributions from the hard thermal
loops for hot QCD are
\cite{Viji},\cite{Braaten90},\cite{Pisarski}
\begin{equation}\label{hard-loops-long}
 \Pi_L(k_0\to 0,K) = \Pi_{00}(k_0\to 0,K) = m_{el}^2 = g^2 T^2
 \left(
 1 + {N_f\over 6} + {1\over 2\pi^2} \sum_f {\mu^2_f \over T^2}
 \right)
\end{equation}
for the longitudinal part and
\begin{equation}\label{hard-loops-trans}
 \Pi_T(k_0\to 0,K) = 0
\end{equation}
for the transverse part of the  gluon polarization tensor.
Substituting Eqs.(\ref{hard-loops-long}),(\ref{hard-loops-trans})
into Eq.(\ref{gluonprop}), one gets
\begin{equation}\label{long-gluon-prop}
 \Delta_L = {1\over k^2-m_{el}^2}, \qquad \Delta_T =
 {1\over k^2}.
\end{equation}

The resummation method for  perturbation theory at finite
temperature as proposed by
Braaten and Pisarski, which resums HTL contributions, works well
if the momentum scale is $gT$.
But in non-abelian gauge theories there is another pernicious
barrier for perturbation theory besides the weak coupling limit:
Static magnetic fields are not screened at the HTL level and
this may lead to a breakdown of perturbation theory at high loop
order. The corresponding infrared singularities are usually
expected to be cured by the dynamical generation of a magnetic
screening mass $m_g (g^2 T)$
\cite{Burgess},\cite{Gullasy},\cite{Thoma},
but the nature of the latter is still unclear.

It is known that damping of collective modes in a QGP occurs.
This dynamical screening, as opposed to static (Debye) screening,
makes it reasonable to consider a damping rate for gluons, that
is to assign them a finite width when studying the collective
behaviour of a QGP. Assuming that the damping rate of the
transverse gluon is $\nu_T$, we can write the transverse gluon
propagator as \cite{Landsman},\cite{Henning}
\begin{equation}\label{gluon-prop2}
 \Delta'_T={1 \over (k_0-i\nu_T)^2-K^2}
\end{equation}
so that the total resummed gluon propagator becomes
\begin{equation}
 G'_{\mu\nu}(k) =
 A_{\mu\nu} (k) \Delta'_T (k) +
 B_{\mu\nu} (k) \Delta_L (k) +
 D_{\mu\nu} (k) \Delta_\xi (k).
\end{equation}
The bare propagator and vertex are given by
\begin{eqnarray}\label{bareprop}
 \Delta_{\mu\nu} (k) &=& -{1 \over k^2}
 \left(
 g_{\mu\nu} - (1-\xi) {k_\mu k_\nu \over k^2}
 \right),\\
 \label{barevert}
 \Gamma_{\mu\nu\rho}(p,q,k) &=& (p-q)_\rho\ g_{\mu\nu} +
 (q-k)_\mu\ g_{\nu\rho} + (k-p)_\nu\ g_{\mu\rho}.
\end{eqnarray}
By making use of Eqs.(\ref{gluon-prop2})--(\ref{barevert}),
one can evaluate the diagram of Fig.\ref{fig.1}(a) and obtain for
 $F$ from Eq.(6) after a rather
lengthy calculation
\begin{equation}\label{selfenergy}
  F = g^2 N_c
  \int{d^D q \over (2\pi)^D} \ \ { 4 \vert P \vert^2 \over (p+q)^2}
  \cdot \left( \Delta_T'(q)-\Delta_L(q)-
  (1-\xi) {( \hat P \cdot \hat Q)^2 \over
  q^2} \right),
\end{equation}
where the hats on the three-vectors $P$ and $Q$ denote unit vectors.

In the following we will apply the formalism of thermal
field dynamics (TFD), since the imaginary part of the ``$1-2$''
component of the Greens function $\Sigma_{12}$ is easier
evaluated in TFD than in the imaginary time formalisme.
By using the well known relation between
the imaginary parts of retarded Greens functions and ``$1-2$''
components of Greens functions in TFD \cite{Fuji},\cite{Kobes}
\begin{equation}\label{S12relation}
 \mbox{Im} \bar \Sigma = -{e^{\beta\vert p_0 \vert} - 1 \over 2
 e^{\beta \vert p_0 \vert/2}} \mbox{Im} \Sigma_{12},
\end{equation}
one immediately obtains
\begin{eqnarray}\label{imaginaryf}
 \mbox{Im} F &=& -g^2 N_c {e^{\beta\vert p_0 \vert} - 1 \over 2
 e^{\beta\vert p_0\vert /2}} \int{d^D q\over (2\pi)^D} \ \  4 \vert P
 \vert ^2\Delta_{12}(p+q)
 \nonumber\\ && \cdot
 \left( \Delta_T^{12}(q)-\Delta_L^{12}(q)-
 (1-\xi) \left( \hat P \cdot \hat Q \right)^2 \Delta_{12}(q)\right).
\end{eqnarray}
The thermal propagators of $1-2$ type have the form \cite{Kobes}
\begin{eqnarray}\label{thermprop12}
 \Delta^{12}_L(q) &=& 2\pi i \delta(q^2-m^2_{el})\cdot n_B(q)
 e^{-\beta\vert q_0\vert /2},
 \\
 \Delta^{12}_T(q) &=& 2\ \mbox{Im} \Delta_T
 \nonumber\\ &=&
 \left[
 { 1 \over (q_0-i\nu_T)^2-Q^2-m_g^2}
 - { 1 \over (q_0+i\nu_T)^2-Q^2-m_g^2}
 \right] \cdot
 n_B(q)e^{-\beta\vert q_0\vert /2}
 \nonumber\\
 &{{\nu_T = 0} \atop \rightarrow}&
 2 \pi i\delta(q^2-m^2_g)\cdot n_B(q)
 e^{-\beta\vert q_0 \vert /2},
 \\
 \Delta_{12}(q) &=& 2 \pi i  \delta(q^2) \cdot n_B (q)
 e^{\beta \vert q_0\vert/2}.
\end{eqnarray}
We included the magnetic mass $m_g$ in the propagators, since both
self--consistent damping and magnetic screening are non--perturbative effects
of 
order $g^2T$ so that one cannot neglect a priori one effect compared to the
other. In addition, including $m_g$ will enable us to
compare our self-consistent results with the previous results,
where the infrared divergence was regularized by a magnetic
mass $m_g$ alone.

Eq.(\ref{thermprop12}) contains the Bose-Einstein distribution
function $n_B$, which
for soft momentum, $q\ll p\sim T$,  reduces to
\begin{equation}\label{bose}
 n_B(q)={1\over e^{\beta \vert q_0\vert }-1}\simeq
 {T\over \vert q_0 \vert }, \qquad
 n_B(p+q)\simeq n_B(p).
\end{equation}
Substituting Eqs.(\ref{thermprop12})-(\ref{bose}) into
Eq.(\ref{imaginaryf}), one can evaluate the integration
by dimensional regularization at finite temperature
\cite{Hou94}. The final result is
\begin{equation}
 \mbox{Im} F = -{g^2 N_cT \vert P \vert \over 4\pi} \left(
 \ln {m_{el}^2 \over m_g^2 + \nu_T^2} + 1 - {2\over\pi}
 \arctan{m_g\over \nu_T}
 +{\cal O}(g^2)\right).
\end{equation}
From the definition Eq.(\ref{gluondamp}) one obtains
\begin{equation}\label{widthequation1}
 \nu_T = -{\mbox{Im} F \over 2 \vert P \vert } =
 {g^2 N_cT \over 8 \pi}\left( \ln {m_{el}^2 \over 
 m_g^2 + \nu_T^2}+1
 -{2\over\pi} \arctan{m_g\over \nu_T}
\right).
\end{equation}

We can rewrite Eq.(\ref{widthequation1}) in terms of a new energy scale
defined by $m_g$  with dimensionless variables $\bar{\nu}_T = \nu_T/m_g$,
 $\bar{T} = T/m_g$, $\bar{\mu}_T = \mu_T/m_g$ and
\begin{equation}
 \bar{m}_{el} = \frac{m_{el}}{m_g} = g \bar{T}
 \left( 1 + \frac{N_f}{6} + \frac{1}{2 \pi}
 \sum_f \frac{\bar{\mu}_f^2}{\bar{T}^2} \right)^{1/2}
\end{equation}
as 
\begin{equation}\label{gluonselfconsistent}
 \bar{\nu}_T =
 {g^2 N_c\bar{T} \over 8 \pi}\left( 2 \ln \bar{m}_{el} -
 \ln(1 + \bar{\nu}_T^2) + 1
 - {2\over\pi} \arctan{1 \over \bar{\nu}_T}
\right).
\end{equation}
This is a self--consistent equation for the rescaled damping rate
$\bar{\nu}_T$ and it is gauge 
independent to leading order in the coupling constant. 

If one neglects the finite width effects for the internal
particles ($\bar{\nu}_T \ll 1$) Eq.(\ref{gluonselfconsistent}) yields
\begin{equation}\label{widthequation4}
 \bar{\nu}_T =  {g^2 N_c \bar{T} \over 8\pi} \ln \bar{m}_{el}^2,
\end{equation}
which recovers the  previous results from
Refs. \cite{Burgess},\cite{Thoma},\cite{Rebhan93}.

If on the other hand selfconsistent damping dominates
magnetic screening ($\bar{\nu}_T \gg 1$) Eq.(\ref{gluonselfconsistent})
reduces to  
\begin{equation}\label{widthequation2}
 \bar{\nu}_T = 
 {g^2 N_c \bar{T} \over 8 \pi}\left( \ln {\bar{m}_{el}^2 \over 
 \bar{\nu}_T^2} + 1
\right),
\end{equation}
which can be rewritten as
\begin{equation}\label{widthequation3}
 \nu_T = 
 {g^2 N_c T \over 8 \pi}\left( \ln {m_{el}^2 \over 
 \nu_T^2} + 1
\right),
\end{equation}
where $m_g$ drops out completely.
This again is a selfconsistent equation for $\nu_T$ and the infrared
divergences encountered in previous calculations are removed 
self--consistently.

The selfconsistent Eqs. (\ref{gluonselfconsistent}) and (\ref{widthequation2})
will prove to have very similar solutions for $\bar{T} \ge 1$, but differ from
the solutions for Eq. (\ref{widthequation4}).

One can calculate the damping rate for a fast quark going
through a QGP at finite temperature and chemical potential
in a way similar to the preceeding calculation for gluons
by evaluating
\begin{equation}\label{dampmu}
 \nu_q=-{1\over 4\vert p\vert} \mbox{Im}\ \mbox{Tr}
 (\slashed p\Sigma) \vert_{p_0=E}.
\end{equation}

Let us consider first the quark self-energy shown in Fig.\ref{fig.2},
where we have included screening and damping rate effects
by means of the effective gluon propagator \cite{Braaten90}.
For hard quarks $(\langle p_q \rangle \sim T)$ it is sufficient
to use the bare quark propagator and bare vertices in the hard
thermal loop approximation.

We will calculate the imaginary part of the self-energy at
finite temperature and chemical  potential using
TFD. For the hard particles one can show that the
main contribution to the interaction rate comes from the soft
momentum transfer range, i.e. $q \sim gT$ in the weak coupling
limit.

Applying the usual Feynman rules of QCD to Fig.\ref{fig.2}, one obtains
\begin{equation}\label{traceeq1}
 \mbox{Tr} (\slashed p \Sigma)\ \equiv H \ = -g^2 C_f \int
 {d^D q \over (2 \pi)^D} {\mbox{Tr} \slashed p \gamma^\mu
 (\slashed p+\slashed q)\gamma^\nu \over (p+q)^2} G_{\mu\nu},
\end{equation}
with  $C_f = {N^2 - 1 \over 2N}$ the Casimir for $SU(N)$.
By using Eqs.(\ref{propagator})-(\ref{ABDfunc}),
Eq.(\ref{traceeq1}) can be cast into the form
\begin{equation}\label{traceeq2}
 H = -g^2 C_f \int
 {d^D q\over (2 \pi)^D} {2 \vert P \vert ^2\over (p+q)^2}
 \left( \Delta'_T(q) - \Delta_L(q) - (1-\xi)
 { (\hat P \cdot \hat Q)^2 \over q^2} \right)
\end{equation}
Then using the hot
propagators in TFD for gluons as given by Eq.(\ref{thermprop12})
and for quarks as follows
\begin{eqnarray}\label{quarkprop}
 i \Delta_{11}(p) = -i \Delta_{22}(p) &=& \slashed p \biggl[
 {1\over p^2 - i \epsilon} - 2 \pi \left(
 \theta(p_0) e^{x_p/2}n_f(x_p) +
 \right. \nonumber\\ && \left.
 \quad \theta(-p_0)e^{-x_p/2}n_f(-x_p) \right) \delta(p^2)\biggr],
 \nonumber\\
 i\Delta_{12}(p) = -i \Delta_{21}(p) &=& -2\pi \slashed p
 e^{-\beta \mu}\biggl[ \theta(p_0) e^{x_p/2}n_f(x_p) +
 \nonumber\\ &&
 \qquad \qquad \quad \theta(-p_0)e^{-x_p/2}n_f(-x_p)\biggr] \delta(p^2),
\end{eqnarray}
where $n_f(x_p) = [e^{x_p}+1]^{-1}$ with $x_p = \beta (p_0+\mu)$
denotes the Fermi-Dirac distribution function, we
obtain
\begin{eqnarray}
\mbox{Im} H &=& - g^2 C_f\int {d^D q\over (2 \pi)^D}
 e^{-\beta\mu/2} 2|P|^2 \sin 2\phi_{p_{0}' + \mu} (2\pi)^2
 \nonumber\\ &&
 \left( \Delta_T^{12}(q)-\Delta_L^{12}(q)-
 (1-\xi)\cos^2 \theta  \Delta_{12}(q)\right).
 \\ \label{sigmap}
 &=&
 4 g^2 C_f T \int {d^{D-1} q\over (2 \pi)^2} \ \  \vert P \vert ^2
 \delta \left( (p+q)^2 \right)
 \nonumber\\ &&
 \qquad \cdot\left( {1\over Q^2+\nu^2}-{1\over Q^2+m^2}-
 (1-\xi){\cos^2\theta \over Q^2}\right).
\end{eqnarray}
Here we used
the following relation between the retarded self-energy
$\bar H$ and the 1-2 component of the self-energy 
$H_{12}$ in TFD
\begin{equation}
 \mbox{Im} \bar H = {e^{\beta\mu/2}\over \sin 2
 \phi_{p_0 + \mu}} \mbox{Im}H_{12},
\end{equation}
where
\begin{eqnarray}
 \cos \phi_{p_0 + \mu} &=& {\theta(p_0) e^{x_p/4} +
 \theta(-p_0)e^{-x_p/4} \over \sqrt{ e^{x_p/2} +
 e^{-x_p/2}}} \quad \mbox{and}
 \\
 \sin \phi_{p_0 + \mu} &=& {\theta(p_0) e^{-x_p/4} +
 \theta(-p_0)e^{x_p/4} \over \sqrt{ e^{x_p/2} +
 e^{-x_p/2}}}.
\end{eqnarray}
Evaluating the integration in Eq.(\ref{sigmap}) by dimensional
regularization at finite temperature \cite{Hou94} we get as
the final result for the width of the quarks
\begin{equation}\label{quarkwidth}
 \nu_q = {g^2 C_f T\over 8\pi} (\ln {m_{el}^2\over m_g^2 + \nu_T^2}
 + 1-{2\over\pi} \arctan{m_g\over \nu_T}).
\end{equation}

Rewritting Eq.(\ref{quarkwidth}) as before in terms of dimensionless variables
$\bar{\nu}_q = $ $\nu_q/m_g$, $\bar{\nu}_T$, $\bar{T}$, $\bar{m}_{el}$ leads to
\begin{equation}\label{finalquarkwidth}
 \bar{\nu}_q = {g^2 C_f \bar{T}\over 8\pi} \left( 
 2 \ln \bar{m}_{el} -
 \ln(1 + \bar{\nu}_T^2) + 1
 - {2\over\pi} \arctan{1 \over \bar{\nu}_T}
 \right),
\end{equation}
which gives $\bar{\nu}_q$ as a function of $\bar{\nu}_T$ and $\bar T$, where
$\bar{\nu}_T$ is determined self-consistently through
Eq.(\ref{gluonselfconsistent}).

Setting the finite width of the internal particle to zero, one
recovers the result given in \cite{Burgess}. 

One thus finds that the damping rates for the quark and gluon
have the same behaviour with temperature and chemical potential
to leading order in the coupling constant
and differ only by a constant factor $C_f/N_c$, which counts the
number of processes available for color relaxation for quarks as
compared to gluons.

\section {Numerical Analysis of Color Conductivity for a QGP}
\label{sec3}

In a QGP the interaction rates or frequencies among particles
should be sensitive to
variations of temperature and baryon density.
Naively one might expect the interaction rates to
increase with increasing temperature 
$T$ or chemical potential 
$\mu$,
but one can see from the self-consistent
Eqs.(\ref{widthequation1}),(\ref{quarkwidth})
that the interaction rates depend strongly only on temperature.
The reason for this is that the chemical potential
$\mu$
reflects only the baryon density in a QGP
and does not describe the color charge density.

We solved numerically the self-consistent
Eqs.(\ref{widthequation1}),(\ref{quarkwidth})
for the gluon and quark  damping rates as functions of temperature for
different values of the chemical potential in the weak coupling
limit ($\alpha_s = 0.1$) for two quark flavors.

Changing back from quantities $\bar \nu$, $\bar T$, where the energy scale is
set by $m_g$ to ordinary temperature $T$ and damping rate $\nu$ we have to
make an ansatz for the magnetic screening mass  $m_g$, which we know to be of
order $g^2 T$: 
\begin{equation}
m_g \equiv c \alpha_s T \qquad \quad \mbox{with} \alpha_s = \frac{g^2}{4\pi}.
\end{equation}

The dependence of Eqs.(\ref{widthequation1}),(\ref{quarkwidth}) on $c$ is
given by a linear decrease of the asymptotic slope with increasing $m_g$ for
fixed $T$, $\Delta \nu_T /\Delta c \vert_{T=const} = - \pi \alpha_s
T = const$. For Eq.(\ref{widthequation4}) on the other hand the change of the
slope of 
$\nu_T$ as a function of $c$ is given by $\Delta \nu_T /\Delta c = - 3/c$, so
that for large $c$ the asymptotic slope does practically not change.

In the following we will adopt a value of $c=3$ as estimated from lattice gauge
theories \cite{degrand82} and classical considerations \cite{biro93}.

For the considered values of $T$ around 200 MeV in Fig. \ref{fig.3} we find an
only slightly flater curve for Eq.(\ref{widthequation1}), where
the IR--singularities were regularized through a combination of selfconsistent
damping and magnetic screening with $m_g = 3 \alpha_s T$ compared to 
Eq.(\ref{widthequation3}), where selfconsistent damping alone was used. 
The latter case (limit $\nu_T \gg 1$) corresponds to $c=0$, that is vanishing
$m_g$ in Eq.(\ref{widthequation1}). 

For Eq.(\ref{widthequation4}) on the other hand, where the IR--singularities
were regularized through a magnetic screening mass alone we find a 
steeper asymptotic slope.


The results as given in Fig. \ref{fig.3} show that, as expected, the
damping rate is not sensitive to the changes
of the chemical potential, and only slightly enhanced with
increasing $\mu$. Otherwise it is approximately proportional
to temperature.

The influence of chemical potential is highest at very low temperature,
corresponding to a cold (and dense) QGP, like for instance the core of a
neutron star.  
For hot and dilute QGP as expected from ultrarelativistic heavy ion collisions 
the influence of chemical potential on the damping rate becomes negligible.

 From the transport theory of the QGP one can derive the static color
 conductivity as given in \cite{Gullasy},\cite{Thoma}:
\begin{equation}
 \sigma_i={\omega_{pl}^2\over \nu}, \quad  \omega_{pl}^2 =  m_{el}^2/3.
\end{equation}
The color conductivity describes the ability of color transport in a QGP. It
is defined as the ratio between the plasma frequency squared and the
interaction rate $\Gamma$ of the particles in a QGP,
where $\Gamma$ is the inverse color relaxation time
and therefore given by the damping rate $\nu$.
Since the interaction rate increases only linearly with temperature
and logarithmically with chemical potential, but the plasma frequency
squared increases quadratically with $T$ as well as $\mu$, color
conductivity is enhanced with increasing $T$ and $\mu$.

As a numerical results for $\sigma$ one obtains the curves
given in Fig.\ref{fig.4} by using the self-consistent solutions for the
damping rates. 
One finds that the quark chemical potential
has some influence on the color conductivity,
especially at low temperatures where our assumption of a
finite chemical potential does apply. 

\section{Discussion and Conclusions}\label{sec4}

In summary, by taking into account the quark chemical potential
and the transverse gluon damping effects, we obtain  the
self-consistent equation for the damping rate of gluons and quarks
in real time finite temperature QCD language, where the infrared
divergence is removed self-consistently 
and the magnetic screening mass leads to a redefinition of the energy scale.

In addition we have solved the equation
for the damping rate numerically and calculated the color conductivity
for a QGP for different chemical potentials with an ansatz for the magnetic
screening mass $m_g$. The results show  that
the chemical potential enhances the color conductivity somewhat
but has practically no effect on the damping rate for gluons and quarks.

Without considering a chemical potential or self-consistent damping
we can recover the results previously derived by other groups.



We conclude that the quark chemical potential has influence on color transport
only at low temperatures and high densities.
Except for low temperatures color conductivity rises linearly with $T$, where
the slope of the curves depends on the mechanisme to regularize the IR
divergences. 
Because of this $T$ dependence
dissipation can not be neglected in
hydrodynamic descriptions of the expansion phase of a hot QGP in
ultrarelativistic heavy ion collisions.

This work is supported partly by the National Natural Science Funds
of China. S.Ochs wants to thank the
Deutsche Forschungsgemeinschaft (DFG),
the Bundesministerium f\"ur
Bildung und Forschung (BMBF), and the Gesellschaft f\"ur
Schwerionenforschung (GSI) for their support.
The authors are indebted to P. Henning for stimulating
discussions and wish to thank Prof. Liu Lianshou and Prof. U.Heinz for their
help. 

\newpage

\section{Figure captions}

\begin{figure}[hb]
\vspace{1.5cm}
\caption{The resummed one-loop gluon self-energy graphs:
   (a) tadpole, (b) gluon loop, (c) fermion loop, (d) ghost loop.
  }\label{fig.1}
\vspace{1.5cm}
\caption{The resummed one-loop quark self-energy graph.
  }\label{fig.2}
\vspace{1.5cm}
\caption{Damping rate $\nu_T$ for gluons versus temperature
   $T$ for different
   quark chemical potentials, with $\alpha_s = 0.1$.
   Dashed lines: without selfconsistent damping for three different chemical
   potentials $\mu =$ 0, 0.5, 1.0 GeV from
   bottom to top.
   Dotted lines: selfconsistent damping without magnetic screening, chemical
   potential $\mu$ as before.
   Solid lines: selfconsistent damping and  magnetic screening, $\mu$ as
   before. 
  }\label{fig.3}
\vspace{1.5cm}
\caption{Color conductivity $\sigma$ versus temperature
   $T$ for
   different quark chemical 
   potentials , with $\alpha_s = 0.1$.
   Coding of lines as in Fig. 3.
  }\label{fig.4}
\end{figure}


\newpage
\begin{thebibliography}{99}

\bibitem{Muller}
  B. M\"uller,  ``Physics of Quark-Gluon-Plasma''
  in: Proceedings of the Workshop on Pre-equilibrium Parton Dynamics,
  ( Ed. Xingnian Wang ), Berkeley, California, 1994;
  J.-P. Blaizot, Nucl. Phys. {\bf A566}, (1994):333c.

\bibitem{Heinz}
  U. Heinz, K. Kajantie and T. Toimela,
  Ann. Phys. {\bf 176}, (1987):218;
  U. Heinz Ann. Phys. {\bf 168}, (1986):148.

\bibitem{Dan}
  P. Danielewicz and M. Gyulassy, Phys. Rev. {\bf D31}, (1985):53,
  Phys. Rev. {\bf D45}, (1992):R17.

\bibitem{Baym}
 G.Baym, et al. Phys. Rev. Lett. {\bf 64}, (1990):1867;
 M.H. Thoma and M. Gyulassy, Nucl. Phys. {\bf B351}, (1991):491.

\bibitem{Burgess}
  C.P. Burgess and A.L. Marini, Phys. Rev. {\bf D45}, (1992):17.

\bibitem{Gullasy}
 A.V. Selikov and M. Gyulassy, Phys. Lett. {\bf B316}, (1993):373.

\bibitem{Thoma}
  M.H. Thoma, Phys. Rev. {\bf D49}, (1994):451,
  Phys. Lett. {\bf B216}, (1991):144.

\bibitem{Viji}
  {H. Vijia and M.H. Thoma, Phys. Lett. {\bf B342}, (1995):212.}

\bibitem{Braaten90}
  {E. Braaten and R.D. Pisarski, Phys. Rev. {\bf D42}, (1990):2156,
   Phys. Rev. Lett. {\bf 64}, (1990):1338.}

\bibitem{Pisarski}
  { R.D. Pisarski, Nucl. Phys. {\bf A525}, (1991):175c,
    Phys. Rev. Lett. {\bf 63}, (1989):1129.}

\bibitem{Landsman}
  {N.P. Landsman and Ch.G. Van Weert,
   Phys. Rep. {\bf 145}, (1987):141.}

\bibitem{Henning}
  {P. Henning, Phys. Rep. {\bf 253}, (1995):325.}

\bibitem{Fuji}
  {Y. Fujimoto, M. Morikawa and M. Sasaki,
   Phys. Rev. {\bf D33}, (1986):590.}

\bibitem{Kobes}
  {R. Kobes, Phys. Rev. {\bf D42}, (1990):562,
   Phys. Rev. {\bf D43}, (1991):1269.}

\bibitem{Hou94}
  {Y. Fujimoto and Hou Defu, Phys. Lett. {\bf B335}, (1994):87.}

\bibitem{Rebhan93}
  {A. Rebhan, Phys. Rev. {\bf D48}, (1993):R3967;
   E. Braaten and A. Nieto, Phys. Rev. Lett. {\bf 73}, (1994):2402.}

\bibitem{degrand82}
  {T.A. deGrand and D. Toussaint, Phys. Rev. D {\bf 25} (1982) 526.}

\bibitem{biro93}
  {T.S. Bir\'o and B. M\"uller, Nucl. Phys. {\bf A561}, (1993) 477.}

\end{thebibliography}
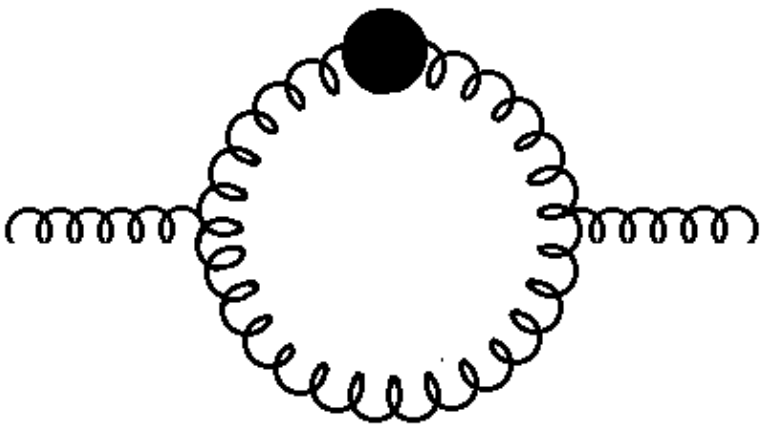
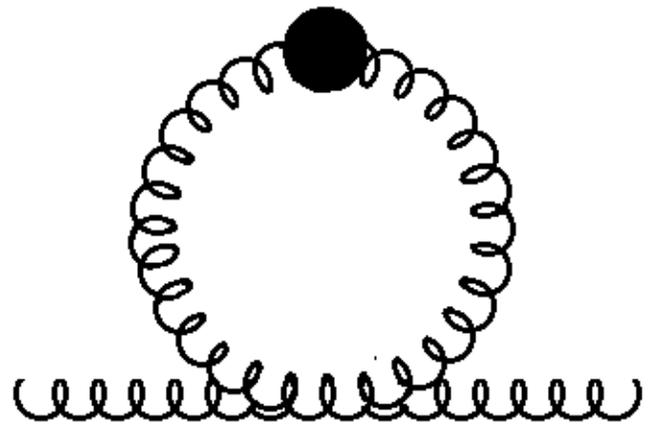
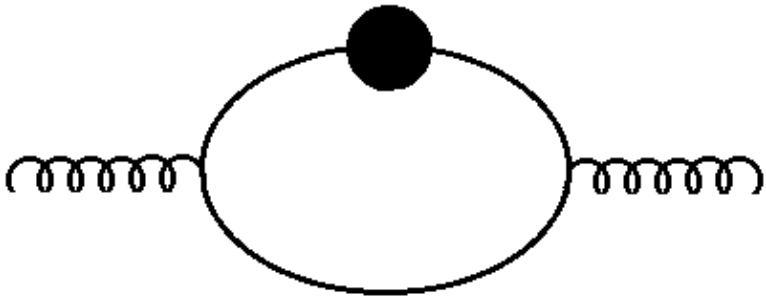
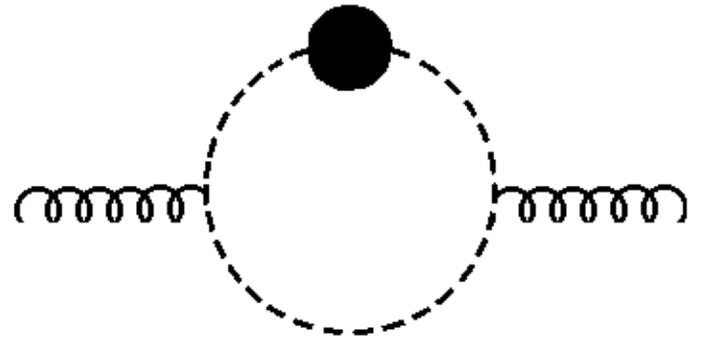
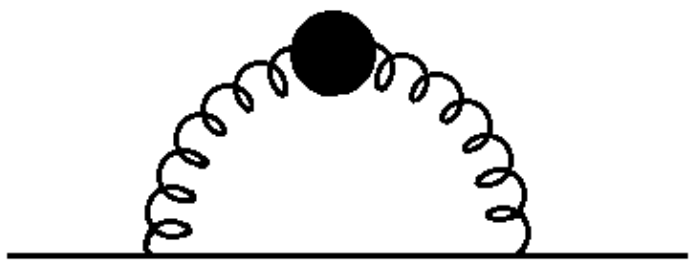
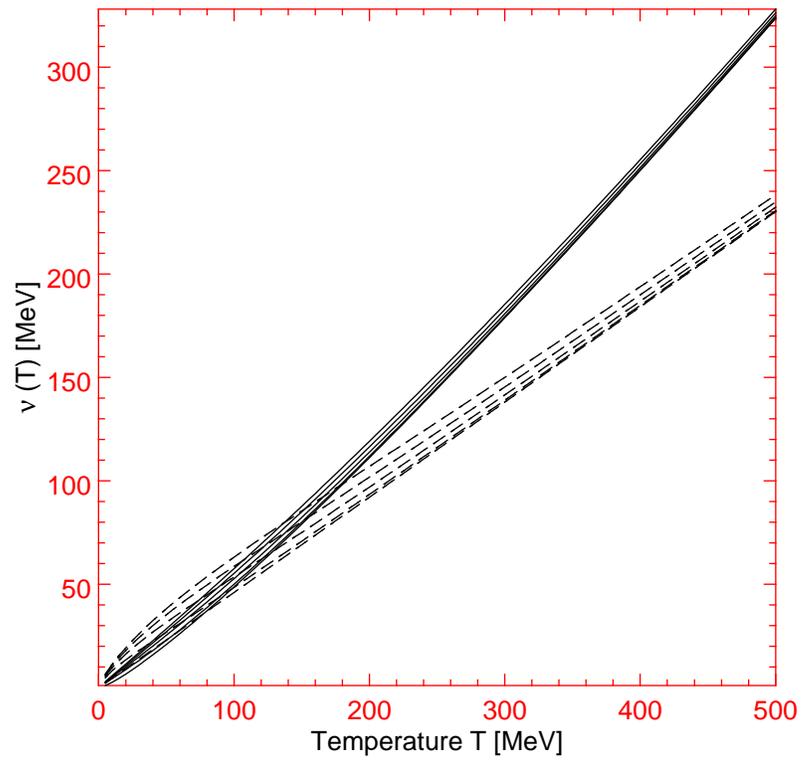
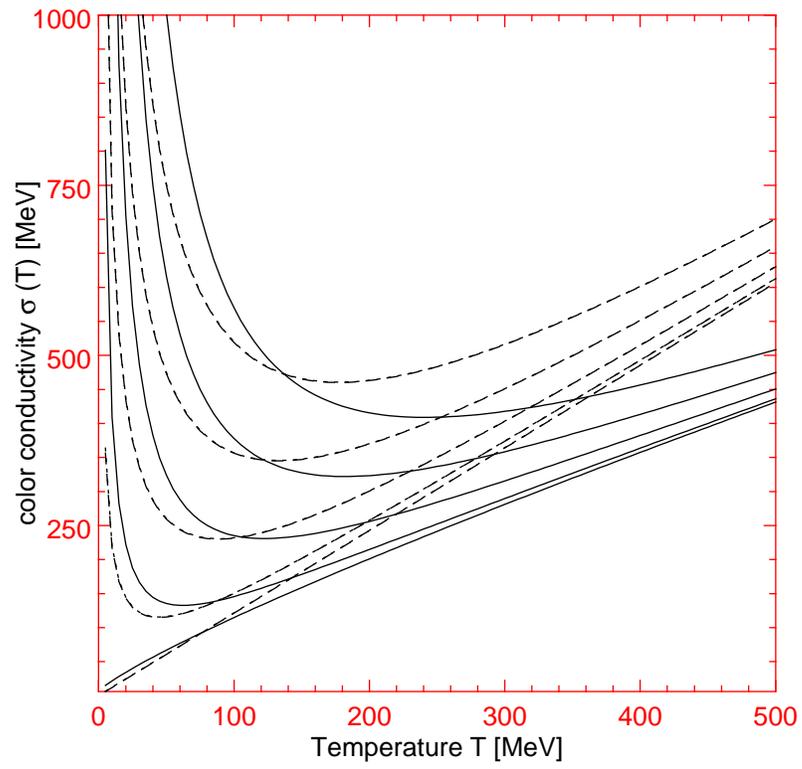
\end{document}